\documentclass[reprint,
    superscriptaddress,
    amsmath,
    amssymb,
    aps,
    prx]{revtex4-2}
    
\usepackage{graphicx}
\usepackage{dcolumn}
\usepackage{bm}
\usepackage{hyperref}
\usepackage[mathlines]{lineno}

\newcommand{\dd}{\mathrm{d}}
\newcommand{\ee}{\mathrm{e}}

\newcommand{\lr}[1]{\left(#1\right)}

\makeatletter
\newcommand*{\rom}[1]{\expandafter\@slowromancap\romannumeral #1@}
\newcommand{\mycomment}[1]{}
\newcommand{\avgcosphi}{\langle \cos\lr{\varphi} \rangle}

\usepackage{comment}
\usepackage{amsfonts}
\usepackage{amssymb,amstext,amsthm,eucal,amsmath}
\usepackage{mathrsfs}
\usepackage{mathtools}
\usepackage{graphicx}
\usepackage{caption}
\usepackage{subcaption}
\usepackage{wrapfig}
\usepackage{array}
\usepackage{float}
\usepackage{enumerate}
\usepackage{enumitem}
\usepackage{upgreek}
\usepackage{listings}
\usepackage{natbib}
\usepackage{physics}
\usepackage{braket}
\usepackage{ragged2e}
\usepackage[utf8]{inputenc}

\usepackage{booktabs}
\usepackage{tabularx}
\usepackage[table]{xcolor}

\usepackage{siunitx}
\usepackage{todonotes}
\usepackage[a4paper]{geometry}
\hypersetup{
    colorlinks,
    linkcolor=[rgb]{0.206756, 0.371758, 0.553117},
    urlcolor=[rgb]{0.120565, 0.596422, 0.543611},
    citecolor=[rgb]{0.8784313725490196, 0.403921568627451, 0.4549019607843137},
    breaklinks=true 
}

\preprint{APS/123-QED}

\begin{document}

\title{\textbf{Observation of area laws in an interacting quantum field simulator}}

\author{Maciej~T.~Jarema}
    \email{ppymj11@nottingham.ac.uk}
    \affiliation{School of Mathematical Sciences, University of Nottingham, University Park, Nottingham, NG7 2RD, UK}
    \affiliation{Centre for the Mathematics and Theoretical Physics of Quantum Non-Equilibrium Systems, University of Nottingham, Nottingham, NG7 2RD, UK}
   \affiliation{Nottingham Centre of Gravity, University of Nottingham, Nottingham, NG7 2RD, UK}
\author{Mohammadamin~Tajik}
    \email{amintajik.physics@gmail.com}
    \affiliation{Vienna Center for Quantum Science and Technology (VCQ), Atominstitut, TU Wien, Vienna, Austria}
    \affiliation{Max Planck Institute of Molecular Cell Biology and Genetics, Dresden, 01307, Germany.}
    \affiliation{Division of Biology and Biological Engineering, California Institute of Technology, Pasadena, CA, USA}
\author{Jörg~Schmiedmayer}
    \affiliation{Vienna Center for Quantum Science and Technology (VCQ), Atominstitut, TU Wien, Vienna, Austria}
\author{Silke~Weinfurtner}
    \affiliation{School of Mathematical Sciences, University of Nottingham, University Park, Nottingham, NG7 2RD, UK}
   \affiliation{Centre for the Mathematics and Theoretical Physics of Quantum Non-Equilibrium Systems, University of Nottingham, Nottingham, NG7 2RD, UK}
   \affiliation{Nottingham Centre of Gravity, University of Nottingham, Nottingham, NG7 2RD, UK}
\author{Tobias~Haas}
    \email{tobias.haas@uni-ulm.de}
    \affiliation{Institut für Theoretische Physik and IQST, Universität Ulm, Albert-Einstein-Allee 11, 89069 Ulm, Germany}
    \affiliation{Centre for Quantum Information and Communication, École polytechnique de Bruxelles, CP 165, Université libre de Bruxelles, 1050 Brussels, Belgium}

\date{\today}


\begin{abstract}
    Information shared between parties quantifies their correlation~\cite{Shannon1948}. The encoding of correlations across space and time characterises the structure, history, and interactions of systems~\cite{Cover2006,Nielsen2010}. One of the most fundamental properties that emerges from studies of information is the area law, which states that information shared between spatial subregions typically scales with the area of their boundary rather than their volume~\cite{Audenaert2002,Hastings2007,Wolf2008}. In non-interacting, quantum many-body systems, where Gaussian statistics apply, the scaling of information measures is well understood~\cite{Calabrese2005,Casini2009,Eisert2010,Langen2013,Tajik2023}. 
    Within interacting systems, the readout of information measures is impeded by the complexity of state reconstruction~\cite{Cramer2010,Lanyon2017}. As such, no measurements beyond small quantum systems (e.g., composed of few, localised particles) have been made~\cite{Islam2015,Kaufman2016,Lukin2019,Brydges2019,Joshi2023}.
    Here, we fill this gap by experimentally demonstrating the area law of mutual information in an ultra-cold atom simulator of quantum fields with tuneable interaction strength~\cite{Schweigler2017}.
    Our results detail the scaling of mutual information with subsystem volume, boundary area, and separation between spatial regions at finite temperature.
    Moreover, we quantify the total effect of non-Gaussian correlations using an information-theoretic measure --- relative entropy.
    Our presented approach is data-driven, model agnostic, and readily applicable to other platforms and observables~\cite{Georgescu2014,Altman2021}, thus constituting a universal toolkit for probing information in high-dimensional quantum systems and its role in shaping quantum matter~\cite{Salzinger2021,Iqbal2024} and spacetime~\cite{Unruh1981, MunozDeNova2019, Haas2022c,Svancara2024}.
\end{abstract}

\maketitle

Physical systems are described by their \textit{state}, which is a set of relevant physical quantities (\textit{degrees of freedom}) at a given time. While the full microscopic state is often complex --- e.g., the positions and velocities of all molecules in a room --- it is usually unnecessary. Instead, systems are frequently characterised by emergent quantities, such as temperature, pressure, or correlation functions. Total correlations are perfectly summarised by measures provided by information theory, which quantify all dependencies within the system of interest. Information revolutionised our understanding of many physical systems, from signal processing in telecommunications~\cite{Cover2006} to cells in microbiology~\cite{Olshausen1996} and quantum effects on microscopic scales~\cite{Nielsen2010}, through the lens of underlying statistical uncertainties, interactions, and organisational patterns.

Among quantum systems, there exists a subclass that produces simple states --- ones that can be fully specified by first- and second-order correlation functions between relevant degrees of freedom. These are referred to as Gaussian states~\cite{Weedbrook2012}.
Using only Gaussian states enables a rich variety of studies and leads to many fundamental and practical applications~\cite{Grosshans2003,Zhong2020}, of which the most relevant for this work is: quantum simulation of free field theories~\cite{Gring2012,MunozDeNova2019,Haas2022c,Haas2025b}.
The emergence of area laws, and information transport within these constraints, are well understood~\cite{Calabrese2005,Casini2009,Eisert2010}, and have been experimentally demonstrated~\cite{Langen2013,Tajik2023,Aimet2025}.

In more complex quantum many-body systems, ones for which Gaussian statistics become insufficient, specifying the state requires higher-order correlation functions between relevant degrees of freedom.
Whilst second and higher-order correlation functions reveal many physical properties of the system~\cite{Schweigler2017}, they are insufficient to tell us about information-theoretic quantities.
The high complexity of generic quantum states renders inferring information measures challenging --- a task deemed impossible for experimental settings that probe continuous quantum fields~\cite{Cramer2010,Lanyon2017}.
This severely limits experimental access to information in the quantum regime, especially beyond small, discrete systems~\cite{Islam2015,Kaufman2016,Lukin2019,Brydges2019,Joshi2023}.
Yet, quantifying the scaling of information while continuous quantum systems interact (producing non-Gaussian states) is necessary, especially in the context of testing area laws in strongly correlated quantum matter~\cite{Lukin2019,Abanin2019,Salzinger2021,Iqbal2024}, certifying systems hard to simulate classically with tensor-network algorithms~\cite{Georgescu2014,Altman2021,Zhong2020}, and the emergence of hydrodynamic descriptions from the underlying correlation structure~\cite{Castro-Alvaredo2016,Cataldini2022}.

Here, we extract information measures from a pair of tunnel-coupled Bose-Einstein condensates --- a continuous, quantum many-body system which we confirm to be firmly within the non-Gaussian regime. By analysing distributions of measured observables, we quantify key features of information, including area laws, correlation decay, and measures of non-Gaussianity, demonstrating their robustness against interactions. Our experimental system has been previously established to simulate the dynamics of interacting quantum fields~\cite{Gritsev2007,Schweigler2017}, allowing us to explore regimes otherwise difficult to access experimentally.

\begin{figure}[]
    \centering
    \hspace*{-0.5cm}
    \resizebox{1.1\linewidth}{!}{
        \includegraphics[
            page=1,
            trim={6.5cm, 19.2cm, 7cm, 4cm},
            clip
        ]{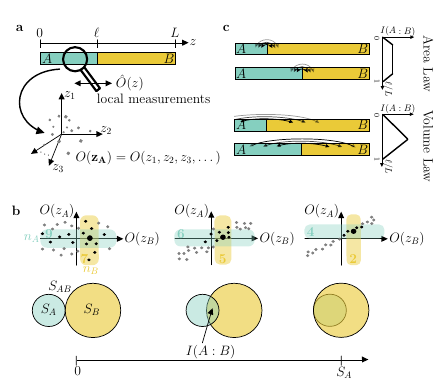}
    }
    \caption{\justifying \small\textbf{Estimating information and its scaling}
    \textbf{a.} A $1$--dimensional system of length $L$ along $z$ is considered to be formed by spatial subregions $A$ ($z\leq\ell$) and $B$ ($z>\ell$). Spatially resolved measurements $\hat{O}\lr{z}$ populate a vector space of collected samples $\hat{O}\lr{\bm{z_A}}$ describing subsystem $A$.
    \textbf{b.} Schematics showing sample correlations in low-dimensional ($D=2$) vector spaces. Top row: patterns formed by mock samples; bottom row: intuitive understanding of shared information. The diagrams range from uncorrelated (left) to fully correlated (right). Within the top row, vertical (orange) and horizontal (green) stripes illustrate $n_A$ and $n_B$ used for computing mutual information as in equation~\eqref{eq:MI_estimator}.  
    \textbf{c.} Diagrammatic representation of information scaling. Top row: area law scaling exemplified by short-range correlations, insensitive to boundary position, resulting in flat scaling of mutual information with subsystem size (right). Bottom row: volume law scaling, dominated by long-range correlations, sensitive to boundary position, resulting in an extensive scaling of mutual information with subsystem size (right).
    }
    \label{fig:Figure_1}
\end{figure}

\begin{figure*}[t!]
    \centering
    \resizebox{\textwidth}{!}{
        \includegraphics[
            page=1,
            trim={2.65cm, 5.15cm, 3cm, 15cm},
            clip
        ]{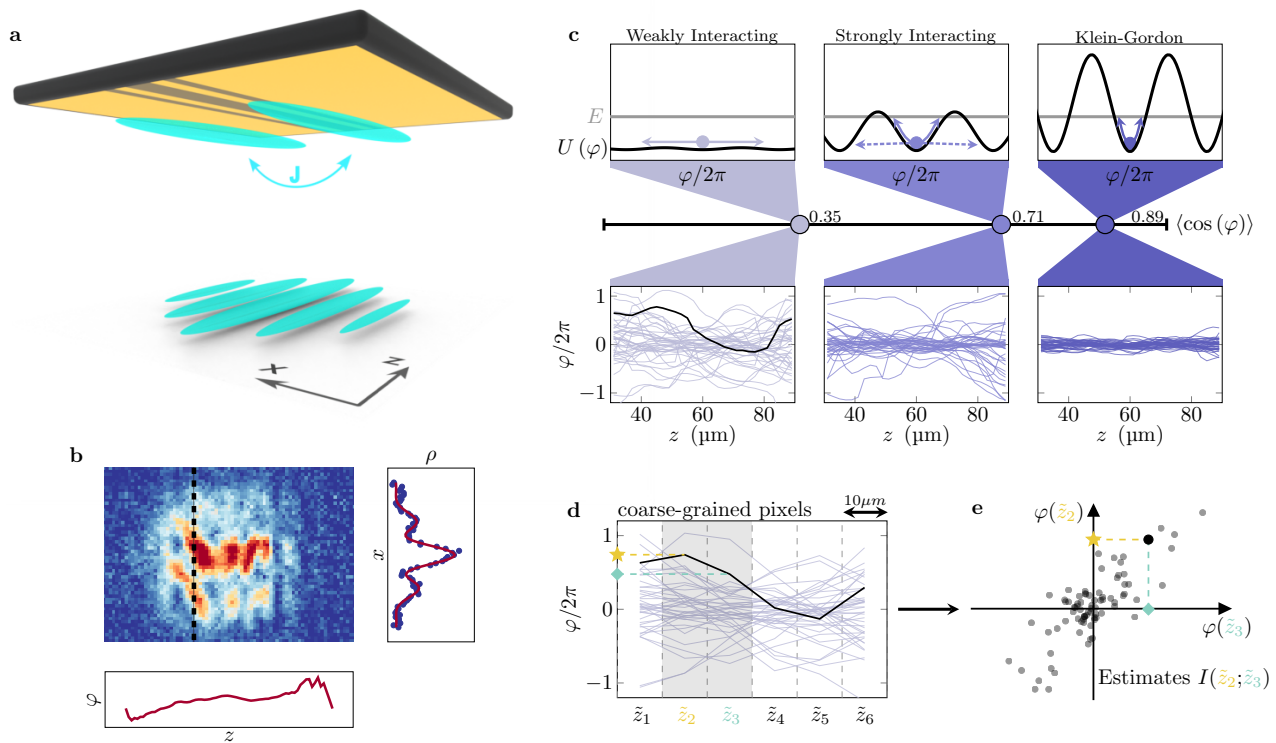}
    }
    \caption{\justifying
        \small\textbf{Model-agnostic extraction of information in an interacting quantum field simulator.}
        \textbf{a.} Schematic of the experiment consisting of two, $1$--dimensional superfluid ${}^{87}\text{Rb}$ condensates, trapped to harmonic potentials under the atom chip and coupled by tunable tunnelling at rate $n_{1\text{D}}J$.
        \textbf{b.} Example phase extraction, showing (left) $2$D projected atomic density, (right) its single $z$ value slice with performed fitting (see methods section), and (bottom) the resulting phase profile over the full $z$ extent. This provides samples of the emergent quantum field being simulated. The illustrated process is repeated $N_s$ times at constant temperature $T$ and coupling strength $J$, to obtain statistics.
        \textbf{c.} Collection of observed phase profiles among various effective interaction regimes. Interaction strength determines phase locking as quantified by the coherence factor $\avgcosphi$. At three representative values of $\avgcosphi$, we plot the experimental phase profiles. The data respects dynamics of: a weakly interacting field (left~:~light purple), a strongly interacting, massive field (middle~:~purple), a non-interacting, massive Klein-Gordon field (right~:~dark purple). Intuition building diagrams of quasi-particle mobility in the potential $U\lr{\varphi}$ (black) at thermal energy $E$ (grey) are provided above. One phase profile is marked out in black.
        \textbf{d.} The spatial profiles are coarse-grained to six pixels labelled by $\tilde{z}_i$. A selection of pixels into two spatial subsystems ($\tilde{z}_2$ and $\tilde{z}_3$) is made by grouping the collected data. The black profile displays the effect of this process; additionally, it serves as an example of data preparation.
        \textbf{e.} Data cloud constructed from two subsystems formed by pixels $\tilde{z}_2$ and $\tilde{z}_3$. The previously outlined black phase profile constructs the data point shown by a black circle, from local values of relative phase marked by a star ($\varphi\lr{\tilde{z}_2}$) and a diamond ($\varphi\lr{\tilde{z}_3}$). Neighbour searches are then used to estimate mutual information between chosen subsystems as shown in Fig.~\ref{fig:Figure_1}.b.
    }
    \label{fig:Figure_2}
\end{figure*}

We begin with local measurements of observable $\hat{\mathcal{O}}\lr{z}$ made on a quantum system (see Fig.~\ref{fig:Figure_1}.a). The results of these measurements are samples $\mathcal{O}\lr{z}$, without the operator hat, from the underlying distribution $f[\mathcal{O}\lr{z}]$, as represented in Fig.~\ref{fig:Figure_1}.a. This distribution is based on the quantum state of the system. We consider a spatial region (or subregion of the full system) $A$, within which the spread of measured values quantifies the subsystem’s uncertainty, expressed in terms of missing information bits.
This can be fully quantified by the Shannon differential entropy $S_A= - \int \dd \mathcal{O}\lr{\bm{z_A}} f [\mathcal{O}\lr{\bm{z_A}}] \ln f [\mathcal{O}\lr{\bm{z_A}}]$. As a visual aid, a broader distribution of points is associated with higher entropy, reflecting the intuitive notion that less information is available about the local system. In Fig.~\ref{fig:Figure_1}.b, the emergence of correlations is reflected in the structure of the joint ($D$ dimensional) distribution: as correlations increase, the initially diffuse cloud of points becomes progressively more elongated, revealing stronger relationships between the observables. Within this setting, mutual information between two subsystems $A$ and $B$ (here considered to be distinct spatial subregions) defined as
\begin{equation}\label{eq:Mutual_Information}
	I(A : B) = S_A + S_B - S_{AB} \,,
\end{equation}
quantifies the amount of information shared between them --- that is, the extent to which their joint statistics differ from those expected from independent subsystems. Owing to this general definition, it captures correlations of arbitrary type as well as order, and when it vanishes, any correlations between the considered regions are ruled out. We estimate mutual information directly from measurement samples by finding distances to the $k^{\text{th}}$ nearest neighbour of each data point $i$, counting the number of points within that separation along each variable axis ($n_{A, i}$, $n_{B, i}$, see Fig.~\ref{fig:Figure_1}.b), and then applying the following estimator~\cite{Kraskov2004}
\begin{equation}\label{eq:MI_estimator}
    I(A: B) \hspace{-0.05cm} = \hspace{-0.05cm} \psi(k) \hspace{-0.05cm} + \hspace{-0.05cm} \psi(N_s) \hspace{-0.05cm} - \hspace{-0.05cm} \frac{1}{N_s} \hspace{-0.05cm} \sum_{i=1}^{N_s}\Big[\psi(n_{A, i}) \hspace{-0.05cm} + \hspace{-0.05cm} \psi(n_{B, i})\Big],
 \end{equation}
where $\psi$ is the digamma function. This estimator provides a formal expression of the intuition, illustrated above, that increased randomness in the sample distribution leads to reduced inter-sample correlation.

Employing this methodology allows us to extract mutual information by analysing experimental samples of relevant degrees of freedom, without reconstructing the underlying distribution or the full quantum state. Such methods were first theoretically proposed in~\cite{Haas2024c,Haas2025a,Haas2025c}, showing that they efficiently capture information and reproduce area law scaling. Thus, we gain access to information-theoretic quantities that help us to experimentally understand complex quantum systems beyond Gaussian approximations.

To demonstrate the power of this approach in terms of characterising information scaling and area law behaviour, we turn to a continuous quantum many-body experiment that simulates quantum fields and allows for precise tuning of higher-order interactions ranging from well-understood Gaussian regimes to far-from-Gaussian scenarios.

In our experiments (Fig.~\ref{fig:Figure_2}.a), we cool and trap two parallel cigar-shaped clouds of ultra-cold ${}^{87}\text{Rb}$ atoms below an atom chip. We use standard techniques of laser- and evaporative cooling to achieve equilibrium temperatures of $20$--$50\, \si{\nano \kelvin}$. A total number of $\approx 10,000$ atoms in two clouds are confined in a double-well trap with an adjustable barrier between them that allows tunnelling of particles at a rate $n_{1\text{D}}J$. Both clouds have the same linear average atomic density of $n_\mathrm{1D} \approx 70\, \si{\micro \meter^{-1}}$ and can be described by a fluctuating bosonic field $\hat{\psi}_{1,2}(z) = \sqrt{n_\mathrm{1D} + \delta\hat{\rho}_{1,2}(z)}e^{i\hat{\theta}_{1,2}(z)}$. Here, $\hat{\theta}_{1}$ and $\hat{\theta}_{2}$ represent the phase fluctuations, and $\delta\hat{\rho}_{1}$ and $\delta\hat{\rho}_{2}$ are the density fluctuations of each cloud.
Successful experimental studies have shown that the dynamics of the relative degrees of freedom of this system, $\hat{\varphi}(z) = \hat{\theta}_1(z) - \hat{\theta}_2(z)$ and $\delta\hat{\rho}(z) = [\delta\hat{\rho}_1(z) - \delta\hat{\rho}_2(z)]/2$, simulate an interacting quantum field theory, namely the sine-Gordon (SG) model~\cite{Gritsev2007,Schweigler2017,Beck2018,Tajik2023}, with the Hamiltonian
\begin{equation}\label{eq:SG-Hamiltonian}
    \hat{H}_{\mathrm{SG}} = \int_0^{L} \mathrm{d}z \Big[ g_\mathrm{1D} \delta\hat{\rho}(z)^2 
    + \frac{\hbar^2 n_\mathrm{1D}}{4m} \lr{\partial_z \hat{\varphi}(z)}^2  + \hat{U} \Big] \,,
\end{equation}
where
\begin{equation}\label{eq:SG U}
    \hat{U} = 2\hbar Jn_\mathrm{1D}\cos\lr{\hat{\varphi}(z)} \,,
\end{equation} is the potential.
Here, $L$ is the physical extension of the cloud, $g_\mathrm{1D}$ the effective $1$D atomic interaction strength, $m$ the mass of ${}^{87}\text{Rb}$ atoms, and finally $\hbar$ the reduced Plank's constant. The relative degrees of freedom follow canonical commutation relations given by $[\hat{\varphi}(z), \delta\hat{\rho}(z')] = i\delta(z-z')$.
Experimental access to a wide range of interaction regimes is achieved by changing the barrier height that sets $J$, and scales $\hat{U}$.

\begin{figure*}[t!]
    \centering
    \hspace*{-0.2cm}
    \resizebox{\textwidth}{!}{
        \includegraphics[
            page=2,
            trim={2cm, 15.5cm, 3cm, 3cm},
            clip
        ]{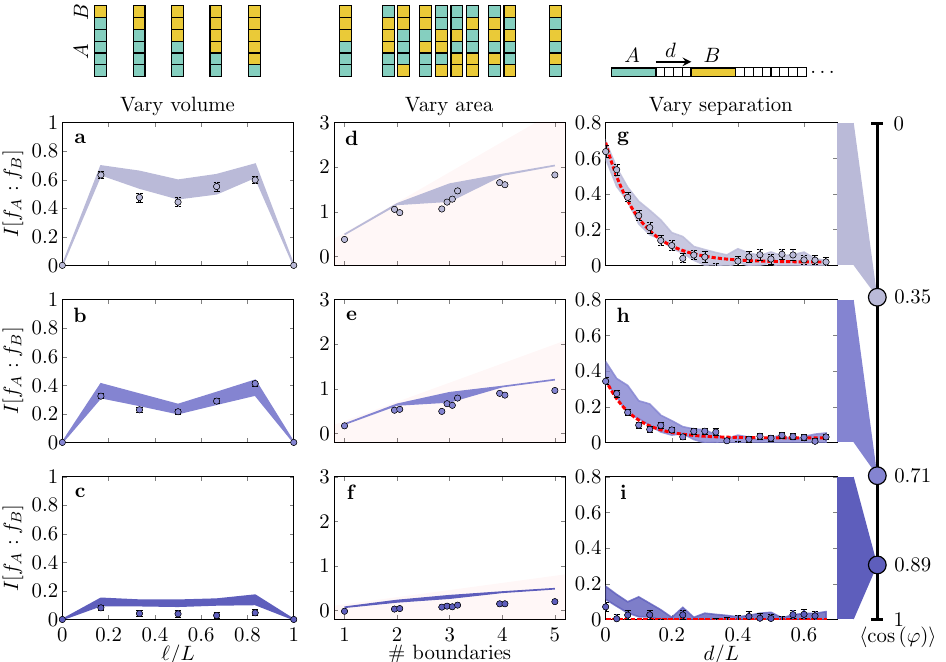}
    }
    \caption{\justifying
    \small\textbf{Area-law behaviour across interaction regimes.} Scaling of information within the same three datasets as in Fig.~\ref{fig:Figure_2} (light purple, purple and dark purple).
    \textbf{a-c}. Mutual information with varying volume; done by displacing the ($0$--dimensional) boundary, as shown above the panels.
    \textbf{d-f}. Mutual information with varying boundary area; done by defining subsystems through disjoint intervals of constant volume ($V_A = V_{B} = 3 \dd \tilde{z}$). Points at different subsystem realisations that yield the same number of boundaries are offset along the $x$-axis for clarity. Additional shading (pink) highlights regions of area law scaling.
    \textbf{g-i}. Mutual information with increasing subsystem separation; done by separating subsystems $A$ and $B$ (size of five $z_i$ pixels each, coarse-grained to $\tilde{z}_{A/B}$). Red lines (dashed) are a guide fit of exponential decay $a e^{-d/\ell_{\text{fit}}} + b$, showcasing the associated correlation length-scale $\ell_{\text{fit}}$.
    In all panels, the errors are estimated using a delete-$d$ Jackknife at $95\%$ confidence intervals. Simulations (shaded regions) are interpolated to match experimental $\avgcosphi$ at a constant $\lambda_T=15 \, \si{\micro \meter}$ and include statistical as well as interpolation error. Additionally, shaded simulations in \textbf{d-f} include the spread over different realisations of the same area. Simulation results are computed using $N_s=2,000$. Both simulations and data analysis use $k=2$ within neighbour search algorithms.}
    \label{fig:Figure_3}
\end{figure*}

The main observable in the experiment is the spatially resolved relative phase, $\varphi(z)$, between the two, tunnel-coupled, $1$--dimensional superfluids. This is the emergent quantity that corresponds to the 'phase quadrature' of the SG quantum field being simulated~\cite{Gritsev2007,Schweigler2017}. We measure the relative phase $\varphi(z)$ using matter-wave interferometry~\cite{Schumm2005}: by switching off all the traps, the atoms fall freely, the two clouds expand in width and overlap. Imaging the resulting interference pattern after a $\approx16\, \si{\milli \s}$ time-of-flight, we obtain $2$D projected atomic densities (Fig.\ref{fig:Figure_2}.a-b). For each slice in the $z$ direction, we extract a single-shot relative phase (Fig.\ref{fig:Figure_2}.b), by fitting a cosine function multiplied by a Gaussian, as detailed in the methods section.

We repeat this protocol at fixed $J$ (fixed $U$) and temperature $T$, measuring an ensemble of $N_\mathrm{s}$ phase profiles $\{ \varphi_i(z_j)|_{U, T} \}$, where $i = 1,2,\dots, N_\mathrm{s}$, and $z_j$ indicates discrete spatial grid of $N_z$ camera pixels. Collecting data within harmonic confinement of the clouds along the $z$ direction results in a low mean density $n_{1D}$ near the edges of the system, which produces unreliable measurements of relative phase in those regions. To mitigate this issue, we only analyse the central $50\%$ of the $z$ extent.
We also compare and contrast the experimental ensembles of relative phase with numerical simulations by employing transfer matrix methods~\cite{Beck2018} (see methods section for more details).

The amplitude of fluctuations in the emergent relative phase field is, on one hand, related to how strongly the two condensates are coupled to each other and, on the other hand, to the temperature --- while increasing the thermal energy $k_\mathrm{B}T$ allows larger relative phase fluctuations, increasing tunnelling suppresses them. The balance between the two energy scales determines the effective coupling strength in the system. This is fully visible within experimental data by the amount of phase locking, which is measured by a coherence factor $\avgcosphi$. It is zero when the relative phase is random, and approaches unity when individual condensate phases lock to an identical value, causing the relative phase to vanish. The wide variety of resulting coupling regimes is demonstrated in Fig.\ref{fig:Figure_2}.c. At low $\avgcosphi$ (low $J$ as compared to $T$ scale), the potential becomes negligible and the quantum system simulates a massless, non-interacting, Gaussian field. At large $\avgcosphi$ (high $J$ as compared to $T$ scale) it simulates a massive, non-interacting, Gaussian (Klein-Gordon) field, and at intermediate $\avgcosphi$ we are dealing with a quantum simulator of a massive, highly non-linear, far-from-Gaussian field that even exhibits topological defects and degenerate vacua (via solitons)~\cite{Schweigler2017}.

\begin{figure}[htbp]
    \centering
    \hspace*{-0.5cm}
    \resizebox{\textwidth}{!}{
        \includegraphics[
            page=3,
            trim={2.5cm, 19cm, 2.5cm, 4.4cm},
            clip
        ]{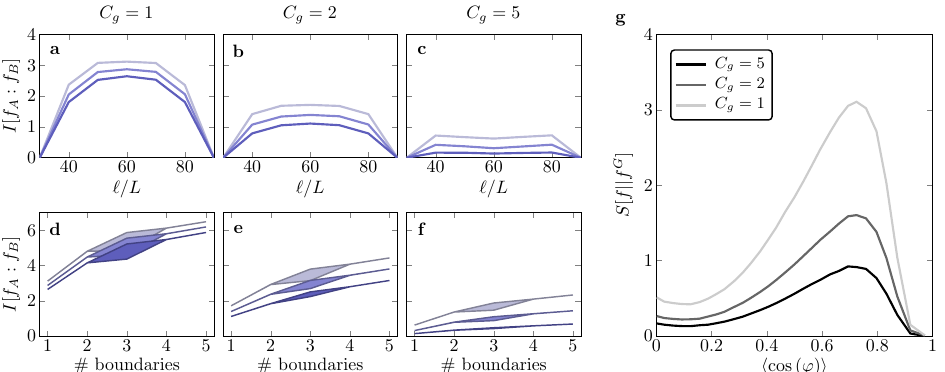}
    }
    \caption{
        \small\justifying\textbf{Effects of Coarse graining.} \textbf{a-c} show the change in the inferred response of mutual information to volume when coarse graining is increased (from no coarse graining -- left, to what is presented in main text -- right). Equivalent analysis in the response of mutual information with area is shown in \textbf{d-f}. Plot \textbf{g} shows the variation of inferred non-Gaussianity behaviour.
        All panels infer the information when less coarse graining is used by simulating the system statistics on poorly resolved spatial grids, keeping all important length scales ($\lambda_T$, $l_J$, $dz$) fixed. The simulations in panels \textbf{a-f} are additionally interpolated onto the three example values of $\langle \cos\lr{\varphi}\rangle$ as in main text and further explained in the methods section. The interpolation errors are not included here.
        }
    \label{ext data fig: Figure_1}
\end{figure}

Collected phase profiles $\varphi_i(z_j)$ are then coarse grained to $N_z=6$ by local averaging, $\varphi_i\lr{\tilde{z}_j} = \frac{1}{|S_j|}\sum_{k\in S_j} \varphi_i\lr{z}_k$, where $S_j$ is a set of original pixel positions of width $|S_j|$ centred at point $z_j$ (see Fig.~\ref{fig:Figure_1}.d). For more information regarding effects of coarse graining on information measures, see the extended data section (Fig~\ref{ext data fig: Figure_1}).
This reduction is required for convergence of the used mutual information estimator~\eqref{eq:MI_estimator}.
The final dataset of relative phase values $\varphi_i(\tilde{z}_j)$ forms our composite system, which can then be spatially sub-selected by grouping the collected data in order to define appropriate subsystems of interest. This is exemplified in Fig.~\ref{fig:Figure_2}.d for two spatial subsystems formed by pixel $\tilde{z}_2$ and pixel $\tilde{z}_3$. Considering various definitions of subsystems allows us to examine the spatial dependence of information, central to studying area laws and correlations within the system.

In order to use equation~\eqref{eq:MI_estimator} and estimate mutual information, we first construct a data cloud on the $D$--dimensional vector space (as indicated in Fig.~\ref{fig:Figure_1}.a-b), where $D$ is the total number of spatial pixels in both subsystems.
An example is provided in Fig.~\ref{fig:Figure_2}.d for a particular phase profile shown in black, producing one point in the data cloud marked by a black circle in Fig.~\ref{fig:Figure_2}.e.

Our data-first analysis directly quantifies mutual information from measurement samples. Our approach avoids bias from fitting collected data or from approximating the underlying distributions.
Additionally, these methods are fully agnostic to geometry, boundaries, model or variable constraints, beyond what is exhibited by the collected data, and hence can be applied to all systems.
On the other hand, the use of an \textit{estimator} requires tests of convergence within available dataset sizes ($N_s$), sensitive to the dimensionality $D$ of data clouds of interest. We show that within available dataset sizes $N_s$, reliable convergence is achieved for dimensions $D\leq 6$ (see supplementary information), requiring our use of coarse-graining techniques described earlier.

Altogether, the above processes (depicted in Fig.~\ref{fig:Figure_1}-Fig.~\ref{fig:Figure_2}) allow us to resolve information structures in the system directly from available experimental data, despite only considering samples of one quadrature of the effective quantum field theory, namely the relative phase $\varphi$ --- not requiring access to the full state of the system. While the full phase space of the emergent quantum field is composed also of $\delta\rho$, we show that in this system, relative phase is sufficient for evidencing the area-law composition of information and characterising non-Gaussianity. The reason behind such strong simplifications is that in the system's Hamiltonian~\eqref{eq:SG-Hamiltonian}, only the phase quadrature exhibits a nonlinearity. Since $\delta\rho$ is decoupled from relative phase, it is Gaussian~\cite{Schweigler2018}.

The spatial encoding of total correlations in our system is revealed through the mutual information~\eqref{eq:Mutual_Information} extracted between various spatial subsystems.
For quantum systems in thermal equilibrium characterised by finite-range interactions (as in this study), equation~\eqref{eq:Mutual_Information} has been predicted to scale at most with the surface area of the boundary between the subsystems rather than their volume --- independent of interaction strength, Gaussianity of the underlying quantum state or the (sub-)system's geometry \cite{Haas2024c}. 
This area law expresses the intuition that the information shared between two spatial regions is concentrated close to their boundary, whereas distant contributions from bulk regions are exponentially suppressed (see Fig.~\ref{fig:Figure_1}.c).
Our experimental findings, presented in the following paragraphs, are consistent with theoretical predictions, revealing clear signatures of area-law behaviour: information does not scale with subsystem volume, grows only sub-linearly with increased boundary area, and decays rapidly with spatial separation of two subsystems. Altogether, we evidence that the information content of the system is governed by boundary geometry rather than bulk volume.

To probe the scaling of the mutual information, we first partition the system with length $L$ into a subsystem of length $\ell$ and the remainder of extent $L-\ell$; then the boundaries' location given by the ratio between left and right $\ell/L$ is varied (as indicated above Fig.~\ref{fig:Figure_3}.a-c). By tuning the interaction strength via the potential $U\lr{\varphi}$ from weak to strong coupling, we systematically investigate the mutual information and find no significant scaling with volume (Fig.~\ref{fig:Figure_3}a-c).
The total correlations reach a plateau that is largely independent of subsystem size, with only deviations caused by coarse-graining effects, in line with the data-processing inequality~\cite{Cover2006,Nielsen2010} (see supplementary information). This confirms that the bulk regions share little to no information, as required by the area law.

We then perform an exhaustive test of the area law by studying the scaling of mutual information with number of boundary points while keeping the subsystem volume fixed --- realised by selection of disjoint partitions as indicated above Fig.\ref{fig:Figure_3}.d-f.
We observe a clear sub-linear scaling, confirming that the composition of spatial correlations follows the area law (shaded regions) across all interaction regimes.

\begin{figure}[h]
    \centering
    \hspace*{-0.5cm}
    \resizebox{\linewidth}{!}{
        \includegraphics[
            page=2,
            trim={6cm, 3.8cm, 7.1cm, 16.3cm},
            clip
        ]{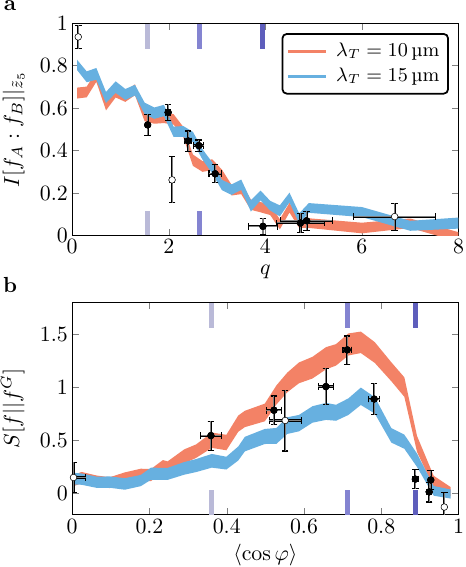}
    }
    \caption{\justifying
    \small\textbf{Interaction strength and temperature dependence of information.}
    \textbf{a.} Decay of mutual information with increasing effective mass parameter $q$. The system is kept at a constant spatial division with $B=\{\tilde{z}_6\}$ and $A$ from the remainder (as indicated on the y-axis label).
    \textbf{b.} Relative entropy measurements of non-Gaussianity as in equation~\eqref{eq:rel_S_of_non_gaussty}.
    In both panels (\textbf{a-b}), shading indicates simulations, with colours specifying the experimental variation with temperature~\cite{Schweigler2017} (low temperature shown in blue, high temperature in red). The points shown in black are experimental results, and empty circles indicate not sufficiently converged experimental results (see extended data section). Axis markers drawn in three shades of purple indicate datasets as in Fig.~\ref{fig:Figure_2} and Fig.~\ref{fig:Figure_3}. Both simulations and data analysis use $k=2$ within neighbour search algorithms.
    }
    \label{fig:Figure_4}
\end{figure}

The aforementioned key physical implication of area laws is the concentration of correlations locally, to boundary regions. It has been shown that a sufficient condition for this to occur is the exponential decay of correlations with distance~\cite{Brandao2013}.
To evidence such decay of correlations and extract characteristic length scales, in Fig.\ref{fig:Figure_3}.g-i, we study the variation of mutual information with increasing subsystem separation. The distance is varied by considering coarse-grained subsystems $A$ and $B$ separated by distance $d$ as indicated above Fig.~\ref{fig:Figure_3}.g-i. Increasing $d/L$, we report exponential decay of correlations from experimental data, in agreement with simulations. The observed, sharp decay of correlations exemplifies further that correlations are insensitive to the behaviour far past the boundary and into the bulk of the subsystem, matching the intuition illustrated in Fig.~\ref{fig:Figure_1}.c. We find that the area-law behaviour persists across the full range of explored interaction strengths and subsystem decompositions. Notably, the non-linear interactions governed by $U$ do not significantly modify this key feature.

Performing fitting using $f_{\text{fit}}(d) = a\ee^{-d/\ell_{\text{fit}}} + b$, we extract the correlation length scale $\ell_{\text{fit}}$ across all interaction regimes and find $\ell_{\text{fit}}$ within the range $3 \si{\micro \meter}$ -- $8 \si{\micro \meter}$, consistent with the expected variation in such systems when excluding non-linear effects~\cite{Tajik2023}.

Having established that mutual information exhibits an area law, we now ask what physical processes govern the magnitude of observed correlations. To do so, we note that the finite correlation length typically arises from the interplay between the mass of the field and the temperature~\cite{Eisert2010}, with larger effective masses and higher temperatures both contributing to shorter correlation length-scales.
To further investigate this connection, we examine the behaviour of mutual information at a constant spatial division of $A$ and $B$ as a function of a parameter $q\propto \sqrt{J}/T$, extractable from experimental data. In the methods section, we show how $q$ can be linearly related to the effective mass in the system at constant temperature. In Fig.~\ref{fig:Figure_4}.a we show the monotonic, exponential decrease of MI with increasing effective mass, across all interaction scales. Note that our results hold despite being surrounded by non-linear interactions in the effective degrees of freedom, whose strength also varies with $q$.
This result not only provides the understanding of the origin of observed local correlations, but also clearly explains a trend in all results presented in Fig.~\ref{fig:Figure_3} --- the overall amount of information decreasing with increasing $\avgcosphi$ due to the associated increase with effective mass.

In Fig.~\ref{fig:Figure_4}.b, we evidence, and fully quantify, the non-Gaussianity of observed states using information-theoretic methods. We do this by considering Shannon relative entropy --- a measure of absolute distinguishability between the true probability distribution $f [\mathcal{O} (\bm{z})]$ and a model distribution $g [\mathcal{O} (\bm{z})]$ \cite{Cover2006,Nielsen2010}

\begin{equation}
    S[f||g] = \int \dd \mathcal{O}(\bm{z}) \, f [\mathcal{O} (\bm{z})] \ln \frac{f [\mathcal{O}(\bm{z})]}{g [\mathcal{O} (\bm{z})]}.
    \label{eq:rel_S_of_non_gaussty}
\end{equation}

\noindent Specifically, we compute relative entropy $S[f||f^G]$ between experimental data (with samples from the underlying distribution $f$), and the \textit{nearest Gaussian} (samples from $f^G$ defined by the same mean and covariance as $f$). $S[f||f^G]$ quantifies how distinguishable the measured data is from being Gaussian~\cite{Cover2006}. This measure can be understood intuitively as the average number of additional information bits required to fully describe the state beyond its Gaussian approximation. Furthermore, it is inversely related to the probability of error in deciding whether the data follows a Gaussian distribution~\cite{Cover2006,Nielsen2010}. 
Relative entropy, therefore, naturally overcomes the fundamental limits of analysing higher-order cumulants, which do not allow for quantitative statements and only indicate non-Gaussianity up to a set order.

We estimate $S[f||f^G]$ directly from experimental samples in line with methods employed to estimate mutual information~\eqref{eq:MI_estimator}. Following~\cite{Wang2009}, we estimate relative entropy from the nearest Gaussian by pseudo-random sampling $N_s$ data points from $f^G$ and utilising neighbour search algorithms. We find the distance to the $k^{\text{th}}$ nearest neighbour of each data point $i$ (defined $\rho^{k}_i$), and its distance to the $k^{\text{th}}$ nearest neighbour within the Gaussian samples (denoted $\nu^{k}_i$), then we construct the estimator:
\begin{equation}\label{eq:kNN_S_estimator}
    S[f||f^G] = \ln \frac{N_s}{N_s - 1} + \frac{D}{N_s}\sum_{i=1}^{N_s}\ln \frac{\nu^{k}_i}{\rho^{k}_i}\,.
\end{equation}

Found relative entropy is shown in Fig.\ref{fig:Figure_4}.b proving the strong non-Gaussian character of the system, in the regime of $\avgcosphi = 0.65-0.8$. The advantage of using information-theoretic methods allows us to directly compare different experimental realisations. Clear comparative statements can now be made, such as ``\textit{Encoding the result of the experiment at $\avgcosphi=0.71$ into Gaussian variables, would require on average $1.7$ bits more data per sample, than when doing so for the experimental results at $\avgcosphi=0.35$}''. As such, this result forms the first quantitative hierarchy of non-Gaussianity in an experiment simulating interacting quantum fields.
Overall, results of Fig.~\ref{fig:Figure_4}.b confirm that conclusions presented in Fig.~\ref{fig:Figure_3} and Fig.~\ref{fig:Figure_4}.a apply within the non-Gaussian, non-linearly interacting regime.

All analyses we perform are readily extendable to other platforms thanks to our \textit{data-first} approach, allowing it to be applied across various architectures~\cite{Georgescu2014,Altman2021}, including ultracold atoms~\cite{Islam2015,Kaufman2016,Haas2022c}, trapped ions~\cite{Lanyon2017,Brydges2019,Joshi2023}, photonics~\cite{Zhong2020,Haas2025b}, and quantum fluids~\cite{Svancara2024} --- as long as a sufficient amount of measurements of relevant quantities can be gathered to resolve desired effects. This directly opens an opportunity to study the scaling of information measures across many regimes, including dynamical and complex non-equilibrium situations. For instance, the breaking of the area law allows for the characterisation of thermalisation or the emergence of many-body localisation~\cite{Abanin2019,Lukin2019}. Further, our methods may be employed to access properties of information transport~\cite{Castro-Alvaredo2016,Aimet2025}. Of particular interest, we would like to highlight the possibility of extending our methods to simulators of quantum fields on curved spacetimes~\cite{MunozDeNova2019,Haas2022c,Svancara2024,Tajik2023b}, which could enable the estimation of the entropy of an analogue black hole horizon and may shed light on large-scale structure formation from quantum fluctuations in the early universe.
When measurements of two non-commuting variables are available~\cite{Kunkel2021,Tajik2023,Pruefer2024}, the considered information measures can also reveal entanglement between spatial regions in strongly correlated regimes~\cite{Haas2024c}.
The work we present finally overcomes the long-standing limitations of information extraction, ultimately enabling its experimental exploration beyond theoretical tractability, in the spirit of quantum simulation.

\let\clearpage\relax
\vspace{-1em}
\bibliography{references.bib}

\section*{ACKNOWLEDGMENTS}\label{acknow_ments} 
We thank the Vienna team for experimental measurements.
We also want to thank Cameron R. D. Bunney, Vitor S. Barroso, for fruitful discussions and feedback. Additionally, we would like to thank Oliver Diekmann for their unmatched help in preparation of Fig.~\ref{fig:Figure_2}.a. MTJ, MT, and SW extend their appreciation to the Science and Technology Facilities Council on Quantum Simulators for Fundamental Physics (ST/T006900/1) as part of the UKRI Quantum Technologies for Fundamental Physics programme. MTJ and SW gratefully acknowledge the support of the Leverhulme Research Leadership Award (RL2019-020). TH acknowledges support from the F.R.S. -- FNRS under project CHEQS within the Excellence of Science (EOS) program and the EU project SPINUS (Grant No. 101135699). SW also acknowledges the Royal Society University Research Fellowship (UF120112,RF/ERE/210198, RGF/EA/180286, RGF/EA/181015) and partial support by the Science and Technology Facilities Council (Theory Consolidated Grant ST/P000703/1). The Vienna team (MT, JS) was supported by the European Research Council: ERC-AdG ``Emergence in Quantum Physics'' (EmQ) under Grant Agreement No. 101097858. 

\section*{Author contributions}
MTJ, TH, and MT wrote the manuscript with contributions from JS and SW. MT, TH and MTJ collaboratively proposed this research project, based on MT's experimental expertise, and TH's earlier findings and developed theoretical methods. MT analysed experimental images and provided relative phase profiles. TH guided the use of information-theoretic quantities, MTJ proposed appropriate estimators, and MT shaped how numerical methods must be applied to experimental data. MTJ performed all further analysis, producing the results of the manuscript and the figures. Scientific guidance was provided by JS and SW. All authors contributed to the discussion of the results. The experiment was conceived by JS.

\section*{Competing interests}
The authors declare no competing interests.

\section*{Data availability}
All analysed data are available from the corresponding author, MTJ. Experimental data is available from JS upon reasonable request.

\raggedbottom
\pagebreak

\section*{Methods}\label{section:Methods}

\noindent\textbf{Cloud preparation and relative phase measurements}

To prepare the experimental system, an atomic cloud of $^{87}$Rb is cooled using standard techniques and magneto-optically guided to the atom chip. On the atom chip, the cloud is further cooled and then confined to the highly elongated geometry shown in Fig.~\ref{fig:Figure_2}.a, producing two cigar-shaped clouds coupled by single-particle tunnelling at a rate $J$. Among the different experimental realisations within our work, tunnelling is altered by adjusting the height of the potential barrier, which determines $J$.

All suspended traps used in these experiments utilised harmonic potentials; thus, the geometry can be fully described by frequencies of each trapping direction as: 
$\omega_{\perp} = 2\pi \times1.4$~kHz (radial confinement) and $\omega_z = 2\pi\times 6.7$~Hz in the (longitudinal confinement). This results in a cloud length of $L=120$~\textmu m, of which the central $60$~\textmu m are analysed, maintaining approximately a constant background density.

After additional evaporative cooling within the double well, reaching temperatures of $T\in[11, 56]$~nK, the measurements are performed. The two resulting clouds contain between $8,000$ and $12,600$ atoms, with a chemical potential $\mu\in 2\pi\hbar[0.70, 0.94]$kHz.

The measurement protocol begins by releasing all spatial trapping, allowing the clouds to free-fall under gravity and expand. After a few milliseconds, this causes matter-wave interference (depicted in Fig.~\ref{fig:Figure_2}.a), which is recorded using absorption imaging after $16$ms (Fig~\ref{fig:Figure_2}.b).
Relative phase at each spatial pixel in the $z$ direction ($\varphi(z)$) is extracted from resulting images by fitting.

\begin{equation}\label{eq:methods:cos_fit}
    f_z(x) \hspace{-0.05cm} = \hspace{-0.05cm} A e^{-\frac{\lr{x-x_0}^2} {\sigma^2_{TOF}}} \hspace{-0.05cm}\left[ \hspace{-0.05cm} 1 \hspace{-0.05cm} + \hspace{-0.05cm} C \hspace{-0.05cm}\cos\lr{2\pi \frac{x-x_0}{\lambda_F} - \varphi} \right] \hspace{-0.05cm} + \hspace{-0.05cm} B,
\end{equation}
where fitting parameters $A, B$ account for amplitude offsets, $C$ determines fringe contrast, $\lambda_F$ sets fringe spacing, $x_0$ and $\sigma_{TOF}$ characterise the centre and width of the Gaussian envelope and $\varphi$ is the relative phase we extract at that chosen position $z$.
This is repeated for all pixels in the $z$ direction ($N_z=60$), producing $\varphi(z)$.
The final result, restricted to $(-\pi, \pi]$, is then phase unwrapped by assuming that $\varphi$ between neighbouring pixels in the $z$ direction does not exceed $2\pi$. Lastly, we select the data for analysis to lay within the central $N_z=30$ pixels.

The final datasets exhibit many striking features relevant to this study.
The spatial average of all phase profiles is non-zero (thus, a zero momentum mode is present), and its measurement distribution strongly changes with coupling strength $J$. As this is the most spatially extended feature, it significantly contributes to all results we present and therefore must be handled with care (see next section).
The experimental relative phase simulates the sine-Gordon field (equation~\eqref{eq:SG-Hamiltonian}), and thus, can exhibit solitons --- high-energy topological kinks that wrap the phase around $2\pi$ in a spatially confined region. These rare events must also be respected within the data analysis (see next section)

\noindent\textbf{Experimental data and estimating information}
We measure information quantities by direct use of relative phase data $\varphi_i\lr{z_j}$, for $i=1, 2 \dots, N_s$ and $j=1, 2, \dots, N_z$ with $N_z=30$, extracted from the experiment as described in the previous section.

Prior to employing equations~\eqref{eq:MI_estimator} and~\eqref{eq:kNN_S_estimator}, the raw data must be appropriately prepared. As mentioned in the main text, the estimation techniques used are fully agnostic to everything except features within the analysed datasets; as such, the data must be ensured to be fully representative of the system's behaviour. Within the current experiment, this is exemplified by the presence of an unphysical redundancy in the collected samples of relative phase, specifically $2\pi$ offsets. 
Whilst, as explained in the previous section, solitons can produce large, local deviations in the relative phase, a uniform offset of the entire spatial profile by multiples of $2\pi$ is an unphysical gauge freedom of the model, experiment and measurement procedure. When processing the data, this arbitrary global shift must be removed in a way that does not disturb any physical features, including the effective energy of the system, zero modes, solitons, or the local structure of the emergent, effective field~\cite{Tajik2023b}.
By restricting the spatial average of relative phase to the unit circle, such that $\langle \varphi_i\lr{\textbf{z}} \rangle_{\textbf{z}} \in \left[0, 2\pi \right], \ \forall i$, we prepare the experimental data in a way that fully preserves all physical properties of the system.

For convergence reasons further specified in the Supplementary Information, we then perform a coarse graining procedure by local, spatial averaging of the data such that $\varphi_i\lr{\tilde{z}_j} = \frac{1}{|S_j|}\sum_{k\in S_j} \varphi_i\lr{z}_k$, where $S_j$ is a set of original pixel positions of width $|S_j|$ centred at point $z_j$. As illustrated in Fig.~\ref{fig:Figure_2}.d, this results in phase profiles $\varphi_i\lr{\tilde{z}_j}$ that lay on a new spatial grid $\tilde{z}_j$ with $N_z=6$, on which local information can be resolved.
From the final dataset, we then compute and store $\avgcosphi$ as well as its $95\%$ confidence interval.

For purposes of computing mutual information between two spatially extended regions $A$ and $B$ (or $A$ and its compliment $A^c$), the final data is then grouped based on positions $\tilde{z}$ that are contained within $A$ or $B$ producing $\varphi\lr{\tilde{\textbf{z}}_A}$, $\varphi\lr{\tilde{\textbf{z}}_B}$. As demonstrated in Fig.~\ref{fig:Figure_3}.d-e, the samples are then positioned on a $D$--dimensional vector space such that each coordinate axis represents a value of relative phase at a single spatial pixel $\tilde{z}_j$. The axes that correspond to relative phase within a selected subsystem $A$ ($B$), altogether form a subspace. For each data point $i$, we then find the distance $\epsilon^{k}_i$ to the $k^{\text{th}}$ (here $2$nd) nearest neighbour.

In order to use the estimator of mutual information~\cite{Kraskov2004} from equation~\eqref{eq:MI_estimator} we then find the number of data points $n_{A}$ ($n_{B}$) that lay within $\epsilon^{k}_i$ of $i$ along the direction of all axes that form subsystem $B$ ($A$). This is demonstrated in Fig.~\ref{fig:Figure_1}.b for the case of $D=2$. When counting $n_A$ and $n_B$, we are careful to exclude any points that lay exactly at a distance $\epsilon^{k}_i$ from point $i$ and include point $i$ itself~\cite{Kraskov2004}.

Estimating relative entropy from the nearest Gaussian follows a similar protocol. For this purpose, we additionally estimate the data mean and variance, then produce $N_s$ samples from the resulting distribution --- the nearest Gaussian. We then perform neighbour searches for each point $i$ within experimental data, finding distances labelled by $\rho^k_i$. Additionally, we estimate distances between each point $i$ within the experimental data and its nearest neighbour within the Gaussian sampled dataset, labelled by $\nu^k_i$. Altogether, using the estimator~\cite{Wang2009} (equation~\eqref{eq:kNN_S_estimator}), we find the relative entropy that is used to quantify total non-Gaussianity.

All computed quantities have been checked for convergence, verified across a range of the free estimation parameter $k$ and their errors were estimated using delete-$d$ Jackknife methods, consistently deleting $5\%$ of the analysed data for each required repetition (checked to not affect the final results and convergence, within experimental and simulation errors). Reported error bars are always stated at $95\%$ confidence interval ($2\sigma$). We note that standard bootstrap methods could not be employed, because the estimators used in equations~\eqref {eq:kNN_S_estimator} and~\eqref{eq:MI_estimator} rely on neighbour searches, and hence, are sensitive to repetition of experimental data points. We have verified that the use of bias correction and acceleration on delete-$d$ Jackknife methods does not alter the results or indicate convergence.

\noindent\textbf{The Sine-Gordon model, simulations and theory}
As mentioned in the main text, the relative degrees of freedom that describe the performed measurements have been successfully shown to follow a sine-Gordon model for a scalar quantum field, and as such, this experiment serves as a simulator of effective quantum fields.
The sine-Gordon model is a description of a one-dimensional, scalar, quantum field with even-power self-interactions mediated by a cosine potential (equation~\eqref{eq:SG U}) at a strength prescribed by a single parameter $J$. As such, in thermal equilibrium, the system is fully specified by only two parameters: $J$ and temperature $T$, or equivalently, their associated length scales $\ell_J^2 = \hbar/\lr{4mJ}$ and $\lambda_T=2\hbar^2 n_{1D}/\lr{m k_{\text{B}} T}$. The length scales ratio $q=\lambda_T/\ell_J$ can be estimated from measurements of $\avgcosphi$ and is used to produce Fig.~\ref{fig:Figure_4}.a. Investigating the expansion of $U$ for low values of $\varphi$ (most appropriate at high $\avgcosphi$ as can be seen from Fig.~\ref{fig:Figure_2}.c) one notices that the dominant term follows $-\hbar n_{1D} J \varphi^2 \sim q^2\varphi^2$, mimicking a field mass when operating at fixed temperature, as alluded to in main text. 

\begin{figure}[H]
    \centering
    \hspace*{-0.5cm}
    \resizebox{0.9\textwidth}{!}{
        \includegraphics[
            page=3,
            trim={2cm, 5.25cm, 3cm, 15.5cm},
            clip
        ]{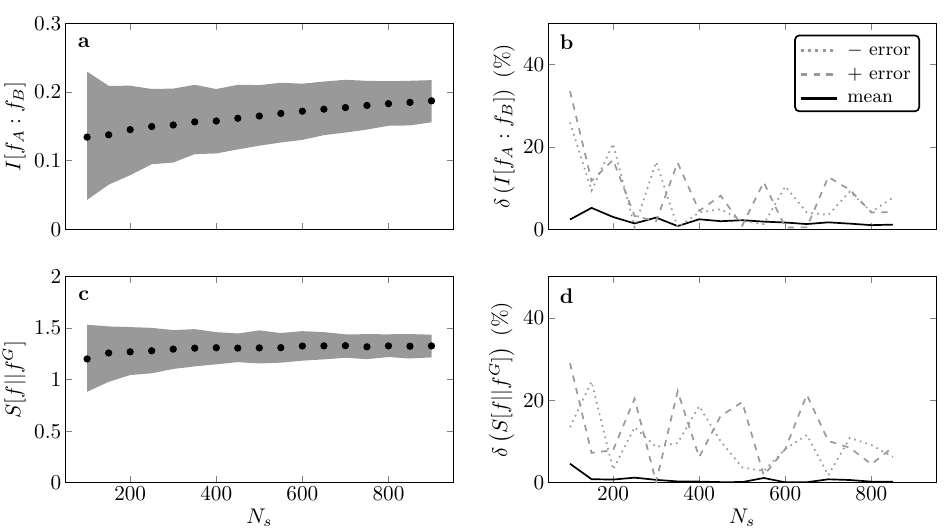}
    }
    \caption{
        \justifying \small\textbf{Convergence of experimental estimates.} \textbf{a-b} variation of estimated mutual information between $A=\{\tilde{z}_1, \tilde{z}_2, \tilde{z}_3\}$ and $B = \{\tilde{z}_4, \tilde{z}_5, \tilde{z}_6\}$ with increasing number of experimental samples. \textbf{a} changes in computed mutual information with delete--$d$ Jackknife errors as shaded regions. \textbf{b} presents percentage convergence error associated with results shown in \textbf{a}.
        \textbf{c-d} variation of estimated relative entropy of non-Gaussianity with increasing number of experimental samples. \textbf{c} changes in computed relative entropy with delete--$d$ Jackknife errors as shaded regions. \textbf{b} presents percentage convergence error associated with \textbf{c}.
        Data for all panels of the above figure comes from a pseudo-random sub-selection of samples from the dataset at index $5$ in table~\ref{tab:data} (middle example throughout the main manuscript). The $N_s$ only extends to $900$ where pseudo-random repetitions correlate and cause error scaling to deviate away from $\propto1/\sqrt{N_s}$. Based on above findings, we consider $N_s \leq 500$ to be not sufficiently converged --- white circles on Fig.~\ref{fig:Figure_4}. 
        }
    \label{ext data fig: Figure_2}
\end{figure}

As first detailed in~\cite{Beck2018}, thermal equilibrium properties of such systems can be simulated by employing powerful transfer matrix techniques. We utilise this approach to simulate statistics of relative phase profiles at constant $\lambda_T$ and $\ell_J$. These are shown as shaded regions in all figures of the main manuscript. All simulations are computed on a spatial grid of $150$ points, interpolated to $60$ (matching experiments) or $\{12, 24, 60\}$ (for coarse-graining studies in Fig.~\ref{ext data fig: Figure_2}). When simulations are compared to experimental findings, we consistently report simulations based on $N_s=2,000$ statistics. When reporting the behaviour of simulations (e.g. Fig.~\ref{ext data fig: Figure_2}), we use fully converged results at $N_s=30,000$, as such, errors are negligible and are thus not shown. In all experimental comparisons, imaging resolution is incorporated into simulations by a Gaussian point-spread function convolution of width $\sigma_{PSF}=3\si{\micro \meter}$, derived from experiment.
As was the case when analysing the data, simulation results are also prepared in the same way (fixing $2\pi$ offsets and coarse graining), and their convergence is checked, further detailed in the Supplementary information. The errors are estimated in the same way, and their convergence is also justified --- see Supplementary Information.

We note that, as detailed above, the simulation relies on being provided the fundamental parameters $q$ and $\lambda_T$, not the phenomenological $\avgcosphi$. As such, in order to compare simulations to experimental data at set $\avgcosphi$ (as done on all results of Fig.~\ref{fig:Figure_3}), we interpolate all simulated information-theoretic quantities and their errors to match experimental values of $\avgcosphi$. We perform this interpolation across a range of values of $q$, at a set $\lambda_T=15\, \si{\micro \meter}$, as informed by experimental temperature estimates ($\lambda_T \in [10, 20]\, \si{\micro \meter}$). The resulting simulation errors include the uncertainty of the experimental value of $\avgcosphi$ as well as the delete-$d$ Jackknife errors, all presented at a $95\%$ confidence interval. This is the reason for simulation errors in Fig.~\ref{fig:Figure_3} being generally larger than those seen in Fig.~\ref{fig:Figure_4}, which only include the delete-$d$ Jackknife considerations.

\onecolumngrid
\raggedbottom
\pagebreak

\section*{Extended data}\label{app:ext_data}

\definecolor{plotcolor1}{RGB}{186, 186, 216}
\definecolor{plotcolor2}{RGB}{132, 132, 209}
\definecolor{plotcolor3}{RGB}{94, 94, 188}
\definecolor{featureColor1}{RGB}{133, 207, 191}
\definecolor{featureColor2}{RGB}{234, 202, 56}

\def\rowW{0.1}

\begin{center}
\captionof{table}{\textbf{Experimental datasets.} Table of all experimental datasets used in this paper. The three experiments exemplified on Fig.~\ref{fig:Figure_2}, Fig.~\ref{fig:Figure_3} and Fig.~\ref{fig:Figure_4} are indicated at indices $1, 5$ and $7$ in the same colour as in main text. $N_s$ specifies the number of experimental repetitions that constitute each dataset and $\avgcosphi$ denotes the average exhibited phase locking.}

\begin{tabularx}{0.8\textwidth}{
    >{\hsize=0.25\hsize}X  
    >{\columncolor{white}}>{\hsize=\rowW\hsize}X  
    >{\columncolor{plotcolor1}}>{\hsize=\rowW\hsize}X  
    >{\hsize=\rowW\hsize}X  
    >{\hsize=\rowW\hsize}X  
    >{\hsize=\rowW\hsize}X  
    >{\columncolor{plotcolor2}}>{\hsize=\rowW\hsize}X  
    >{\hsize=\rowW\hsize}X  
    >{\columncolor{plotcolor3}}>{\hsize=\rowW\hsize}X  
    >{\hsize=\rowW\hsize}X  
    >{\hsize=\rowW\hsize}X  
    >{\hsize=\rowW\hsize}X  
}

\toprule
\textbf{Index} & 0 & 1 & 2 & 3 & 4 & 5 & 6 & 7 & 8 & 9 & 10 \\
\midrule
\textbf{$\avgcosphi$} & $0.00$ & $0.35$ & $0.52$ & $0.55$ & $0.66$ & $0.71$ & $0.78$ & $0.89$ & $0.92$ & $0.93$ & $0.96$ \\
\textbf{$N_s$} & $460$ & $620$ & $901$ & $187$ & $658$ & $1800$ & $774$ & $735$ & $525$ & $598$ & $225$ \\
\bottomrule
\label{tab:data}
\end{tabularx}

\end{center}

\section{All dataset area laws}

\begin{figure}[H]
    \centering
    \resizebox{0.6\textwidth}{!}{
        \includegraphics[
            page=6,
            trim={2cm, 1.8cm, 3.5cm, 2.4cm},
            clip
        ]{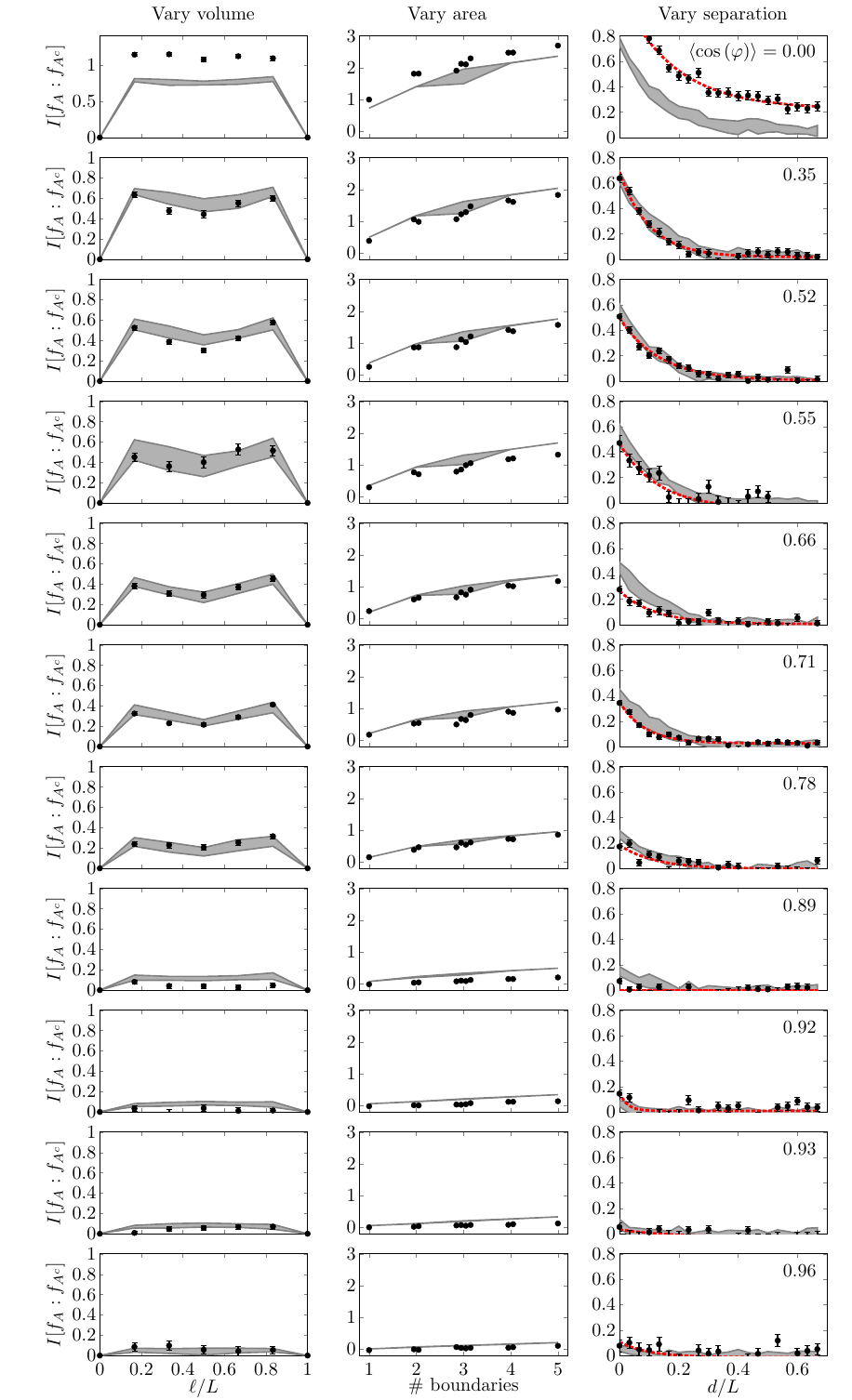}
    }
    \caption{\justifying
        \small\textbf{Extended datasets.} Same analysis and presentation as Fig.~\ref{fig:Figure_3}, extended to all $11$ collected datasets. Note, the $0$th dataset suffers from low data sample size, and poor interpolation of simulations as it lies at the edge of the coherence parameter's range, i.e., $\langle\cos\lr{\varphi}\rangle\approx0.00$. Compare to Fig.~\ref{fig:Figure_4}.a. 
        }
    \label{ext data fig: Figure_ext_MI_AVd}
\end{figure}

\section{Effects of coarse graining and details behind further data treatment}

\section{Numerical stability and data requirements} 

\section*{supplementary information}\label{app:1}

\section{Sensitivity of information to all experimentally characteristic length scales}

Extending our approach of learning about the system behaviour through the use of information-theoretic quantities, in this section, we report the response of information to varying characteristic length scales of the system.

\begin{figure}[H]
    \centering
    \hspace*{-0.5cm}
    \resizebox{\textwidth}{!}{
        \includegraphics[
            page=5,
            trim={2.5cm, 13cm, 3cm, 12cm},
            clip
        ]{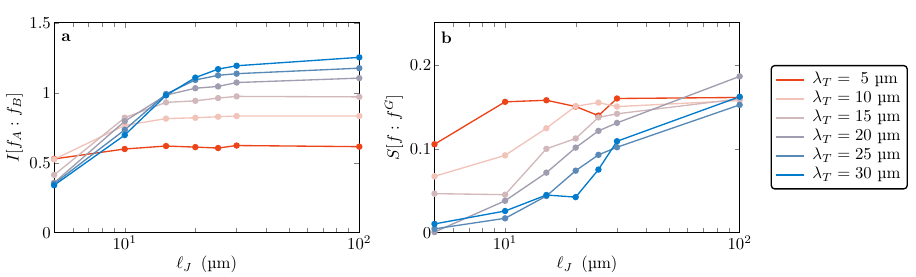}
    }
    \caption{
        \small\justifying\textbf{Response of information to changes in length scales} \textbf{a} Changes of mutual information with varying coupling (coherence) length scale $\ell_J$, for a range of thermal length scales $\lambda_T$. The spatial configuration is chosen as a symmetric split (such that $\tilde{\bm{z}}_A = \{\tilde{z}_1, \tilde{z}_2, \tilde{z}_3\}$ and $\tilde{\bm{z}}_B = \tilde{\bm{z}}_{A^c} = \{\tilde{z}_4, \tilde{z}_5, \tilde{z}_6\}$).
        \textbf{b} Changes of relative entropy of non-Gaussianity with varying coupling (coherence) length scale $\ell_J$, for a range of thermal length scales $\lambda_T$.
        In both panels, simulations are produced with $N_s=30,000, dz=2\si{\micro \meter}$. Coarse-graining effects are not included in the same way as for Fig.~\ref{ext data fig: Figure_1}. The colours of values of $\lambda_T$ are chosen such that red corresponds to higher temperatures (low $\lambda_T$) and blue reflects lower temperatures (high $\lambda_T$). 
        }
    \label{SI fig: Figure_4}
\end{figure}

Fig.~\ref{SI fig: Figure_4}.a shows mutual information between left and right components of the system (half-half split) with decreasing coupling strength $J$ (increasing $\ell_J$) at a range of selected thermal length scales $\lambda_T$ (from hot - red, to cold - blue).
MI is largely insensitive to temperature (slowly increasing) at low $\ell_J$ (strong coupling), and highly sensitive at high $\ell_J$ (weak coupling). The opposite is also true --- MI is only sensitive to changes in $\ell_J$ at low temperature. Overall, we conclude that information only responds to one significant length scale in the system, and hence can be well exemplified by $\avgcosphi$ or $q$.
Furthermore, MI decreases with temperature when $\ell_J$ is large (i.e., weak coupling). 

The response of relative entropy (Fig.~\ref{SI fig: Figure_4}.b) presents essentially the same conclusions: non-Gaussianity is most sensitive to one dominant length scale and hence, can be well exemplified by $\avgcosphi$ or $q$. However, crucially, it only varies highly with temperature when $\ell_J$ is small (strong coupling) and is insensitive to temperature when $\ell_J$ is large (weak coupling). The contrary is also true: high temperature curves are least sensitive to changes in $\ell_J$. We also note that non-Gaussianity increases with increasing temperature at low $\ell_J$ (strong coupling).

\section{Simulation convergence}

\begin{figure}[H]
    \centering
    \resizebox{0.8\textwidth}{!}{
        \includegraphics[
            page=4,
            trim={2cm, 17.2cm, 3cm, 3.5cm},
            clip
        ]{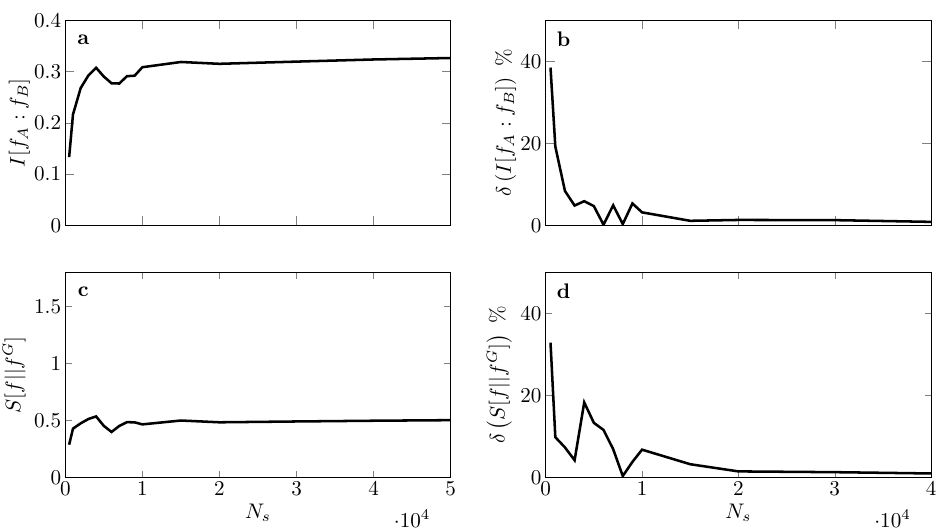}
    }
    \caption{
        \small\justifying\textbf{Convergence of simulated estimates.} \textbf{a-b} variation of estimated mutual information between $A=\{\tilde{z}_1, \tilde{z}_2, \tilde{z}_3\}$ and $B = \{\tilde{z}_4, \tilde{z}_5, \tilde{z}_6\}$ with increasing number of simulated samples. While \textbf{a} shows the changes in computed mutual information, \textbf{b} presents their associated percentage convergence error.
        \textbf{c-d} show variation of the estimated relative entropy of non-Gaussianity with increasing number of simulated samples. While \textbf{c} shows the changes in computed relative entropy, \textbf{b} presents the associated percentage convergence error.
        All plots are kept without interpolation to any experimental $\langle \cos\lr{\varphi}\rangle$, both assuming $q=2, \lambda_T=15\si{\micro \meter}$, which results in converged $\langle \cos\lr{\varphi}\rangle\approx0.48$.
        }
    \label{SI fig: Figure_1}
\end{figure}

Throughout the main manuscript, alongside the results from experimental data, we provided simulations at $N_s=2000$ that resemble realistic experimental requirements. Here, we show that, provided sufficiently many samples, all quantities converge to a determined value --- ruling out any ill-behaved results.
In Fig.~\ref{SI fig: Figure_1}, we show that both mutual information and relative entropy are well behaved (plateau at high $N_s$) and their error can be trusted to $\lesssim 2\%$ accuracy.

\begin{figure}[H]
    \centering
    \resizebox{0.9\textwidth}{!}{
        \includegraphics[
            page=4,
            trim={2cm, 4.5cm, 3cm, 15.5cm},
            clip
        ]{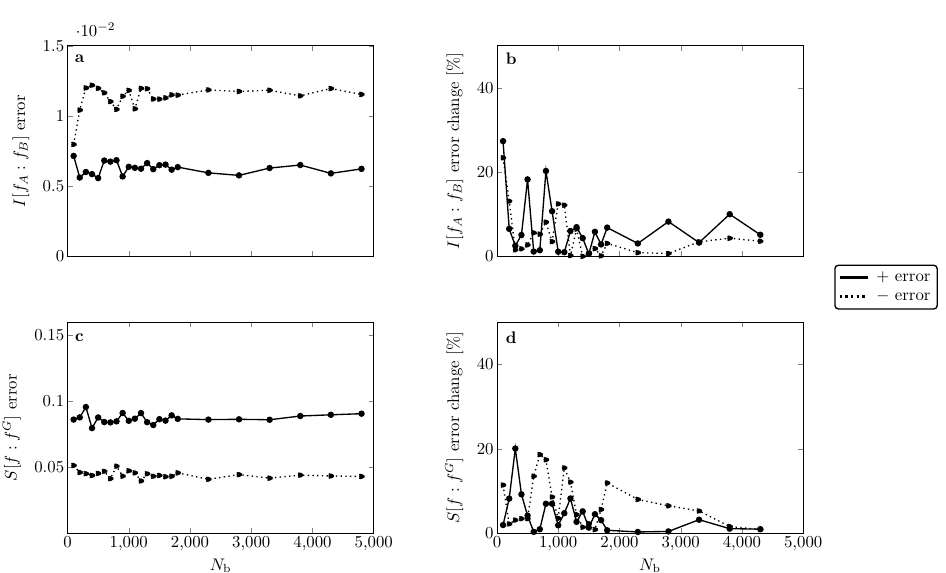}
    }
    \caption{
        \small\justifying\textbf{Convergence of error estimation} \textbf{a} Error estimate on mutual information with increasing number of delete--$d$ Jackknife repetitions. \textbf{b} Percentage change in the estimated error on mutual information (see \textbf{a}) showing convergence.
        \textbf{c} Error estimate on relative entropy of non-Gaussianity with increasing number of delete--$d$ Jackknife repetitions. \textbf{d} Percentage change in the estimated error on relative entropy (see \textbf{c}).
        For both \textbf{a} and \textbf{b}, the spatial configuration is chosen as a symmetric split (such that $\tilde{\bm{z}}_A = \{\tilde{z}_1, \tilde{z}_2, \tilde{z}_3\}$ and $\tilde{\bm{z}}_B = \tilde{\bm{z}}_{A^c} = \{\tilde{z}_4, \tilde{z}_5, \tilde{z}_6\}$).
        All panels use the dataset at index $5$ in Tab.~\ref{tab:data} at $\langle \cos\lr{\varphi} \rangle\approx 0.71$ (middle example throughout the main manuscript).
        }
    \label{SI fig: Figure_2}
\end{figure}

Similarly, in Fig.~\ref{SI fig: Figure_2}, we evidence convergence of errors estimated through delete-$d$ Jackknife methods at $5\%$ removal. Beyond evidencing the well-behaved nature of our estimation, we observe that the number of repetitions used ($3000$) is sufficient to trust the reported error within $\lesssim 10\%$ accuracy.

\end{document}